\documentclass{iaus}
\usepackage{graphicx}

\title[Stellar black holes] %% give here short title %%
{Stellar black holes: Cosmic history and feedback at the dawn of the
universe }

\author[I.F. Mirabel]   %% give here short author list %%
{I. Felix Mirabel $^{1,2}$}

\affiliation{$^1$
CEA-Saclay, IRFU/DSM/Service d'Astrophysique. 91191 Gif sur Yvette. France
\\[\affilskip]
$^2$IAFE-UBA-CONICET. cc 67, suc. 28. (C1428) Buenos Aires. Argentina
\\[\affilskip] email: {\tt felix.mirabel@cea.fr}}
\pubyear{2010}
\volume{275}  %% insert here IAU Symposium No.
\pagerange{xxx}
\jname{Jets at all Scales}
\editors{G.~E. Romero, R.~A. Sunyaev \& T. Belloni, eds.}
\begin{document}
\maketitle
\begin{abstract}
Significant historic cosmic evolution for the formation rate of
stellar black holes is inferred from current theoretical models of
the evolution of massive stars, the multiple observations of compact
stellar remnants in the near and distant universe, and the cosmic
chemical evolution. The mean mass of stellar black holes, the
fraction of black holes/neutron stars, and the fraction of black
hole high mass X-ray binaries (BH-HMXBs)/solitary black holes
increase with redshift. The energetic feedback from large
populations of BH-HMXBs form in the first generations of star burst
galaxies has been overlooked in most cosmological models of the
reionization epoch of the universe. The powerful radiation, jets,
and winds from BH-HMXBs heat the intergalactic medium over large
volumes of space and keep it ionized until AGN take over. It is
concluded that stellar black holes constrained the properties of the
faintest galaxies at high redshifts. I present here the theoretical
and observational grounds for the historic cosmic evolution of
stellar black holes. Detailed calculations on their cosmic impact
are presented elsewhere (Mirabel, Dijkstra, Laurent, Loeb, \& Pritchard  2011). 
\end{abstract}

\firstsection % if your document starts with a section,
\section{The dark ages}

Motivated by a talk of Rashid Sunyaev at the 7th Microquasar
Workshop, I became interested in exploring the possible role of
black holes of stellar mass in the early cosmic evolution, and in
particular, on how feedback from accretion of the remnants of
massive stars could have affected the intergalactic medium during
the dark ages. The so called ``dark ages" of the universe began
about 400,000 years after the Big Bang as matter cooled down and
space became filled with neutral hydrogen for hundreds of millions
years. How most of the matter in the universe became again ionized
(reionized) in less than a billon year is a question of topical
interest in cosmology. The recent detection of the explosion of a
massive star (Salvaterra R. et al. 2009) at $z \sim 8.2$ and the
observation of galaxies (Bouwens et al. 2010) up to $z \sim 8$
support the idea that the ultraviolet radiation from massive stars
in the first galaxies played an important role in the reionization.
However, from the luminosity density of the most distant galaxies it
has been claimed that the UV flux available from massive stars may
not have been enough to keep fully reionized the universe (Bouwens
et al. 2010). Because X-rays have a longer mean free path than the
ultraviolet photons, it is proposed that accreting stellar black
holes provided heating and secondary ionizations over large volumes
of space (Mirabel et al. 2011).

%\begin{figure}[tbd]
%\center
%\includegraphics[angle=270,width=0.6\textwidth]{vbfig1b_v2.eps}
%\caption{
%Computed spectral energy distribution for a typical HMMQ (see Bosch-Ramon et al. 2006a).}
%\label{hm}
%\end{figure}

\section{Formation of black holes by implosion: cosmic historic evolution of BH-HMXBs}
Theoretical models show that the evolution and final fate of massive
stars strongly depend on the initial metallicity and rotation. Stars
with low metal content and initial masses of a few tens of solar
masses collapse directly as black holes, with no energetic supernova
natal kicks (Heger et al. 2003; Meynet \& Maeder 2005). On the other
hand, recent hydrodynamic simulations of the formation of stars with
low metal content (Krumholz et al. 2009; Turk et al. 2009, Stacy et
al. 2009) show that a substantial fraction of these stars form as
small multiple systems dominated by binaries with typical masses of
tens of solar masses. Therefore, from current theoretical models it
is inferred that the majority of  high mass stellar binaries of low
metallicity should remain gravitationally bound after the implosion,
ending as BH-HMXBs, which are known to be powerful sources of
X-rays, massive winds, and relativistic jets (microquasars, see
Mirabel \& Rodr\'\i guez 1999). In the context of these models and
the cosmic evolution of metallicity and star formation evolution it
is then expected that: \textit {1) the mass of stellar black holes,
2) the fraction of black holes/neutron stars, and 3) the fraction of
black hole binaries/solitary black holes should increase with
redshift. Therefore, the formation rate of BH-HMXBs must have been
significantly larger in the early universe than in later epochs.}

\vspace{3mm}
\textbf{Formation rate of stellar black holes as function metallicity}
\vspace{1mm}

How massive stars evolve and die depends on their initial mass,
metal content, angular momentum, and whether they are born in
isolation or in multiple systems (Meynet \& Maeder 2005). Despite
these complexities the expected cosmic evolution of BH-HMXBs
mentioned above is consistent with the following observations of
stellar black holes and neutron stars in the near and distant
universe.

a) In agreement with the theoretical expectations (Heger et al.
2003, Meynet \& Maeder 2005), the masses of black holes in high mass
x-ray binaries determined dynamically seem to be a decreasing
function of the metallicity of the host galaxy (Crowther et al.
2010). The black holes in the high mass binaries M 33 X-7, NGC 300
X-1 and IC10 X-1 which are in small galaxies of low metallicity,
have masses in the range of 16 to 30 solar masses, which are larger
than the mass of any known stellar black hole in the Milky Way and
Andromeda galaxies.

b) It has been proposed that most ultraluminous X-ray sources (ULXs)
are BH-HMXBs that contain black holes of several tens of solar
masses in low metallicity environments, accreting in a slightly
critical regime (Zamperi \& Roberts 2009). It is found that the
occurrence rate of ULXs per unit galaxy mass in nearby galaxies is a
decreasing function of the mass of the host galaxy (Swartz et al.
2009), namely, of its metal content. One extreme case in the local
universe is that of the metal-poor ($Z < 0.05 Z_\odot$) ring
Cartwheel galaxy, where it is estimated (Mapelli et al. 2009) that
more than $\sim 100$ stellar black holes of $30-80$ solar masses
might have been generated via direct collapse (implosion) during the
last $10^7$ yr. The X-ray luminosity of this small galaxy of low
metallicity is $10^{42} {\rm erg}/{\rm s}$, which rivals that of low
luminosity AGN.

c) The kinematics in three dimensions of Galactic black hole
binaries relative to their birth place provides evidence for black
hole formation by implosion. So far, the space kinematics has been
determined for five Galactic black hole binaries. The kinematics of
the three black holes that have masses equal or larger than $10$
solar masses suggest that they have been form directly, without very
energetic supernovae, whereas the two black holes with less than
$10$ solar masses were form with energetic natal explosions. In
fact, the microquasars GRO J1655-40 and XTE 1118+48 which host
respectively black holes of $5$ and $7$ solar masses have runaway
space velocities relative to their environment of $112\pm18 \ {\rm
km}/ {\rm s}$ and $177\pm33 {\rm km}/{\rm s}$, respectively (Mirabel
et al. 2002, Mirabel et al. 2001). On the contrary, Cyg X-1 which
contains a black hole of $\sim 10$ solar masses remained anchored at
its birth place in the association Cyg OB3, and not more than one
solar mass could have been suddenly ejected during a possible natal
supernova (Mirabel \& Rodrigues, 2003).

\begin{figure*}
\center          %    l   b   r   t
\includegraphics[trim=0mm 0mm 0mm 0mm, clip,width=0.8\textwidth]{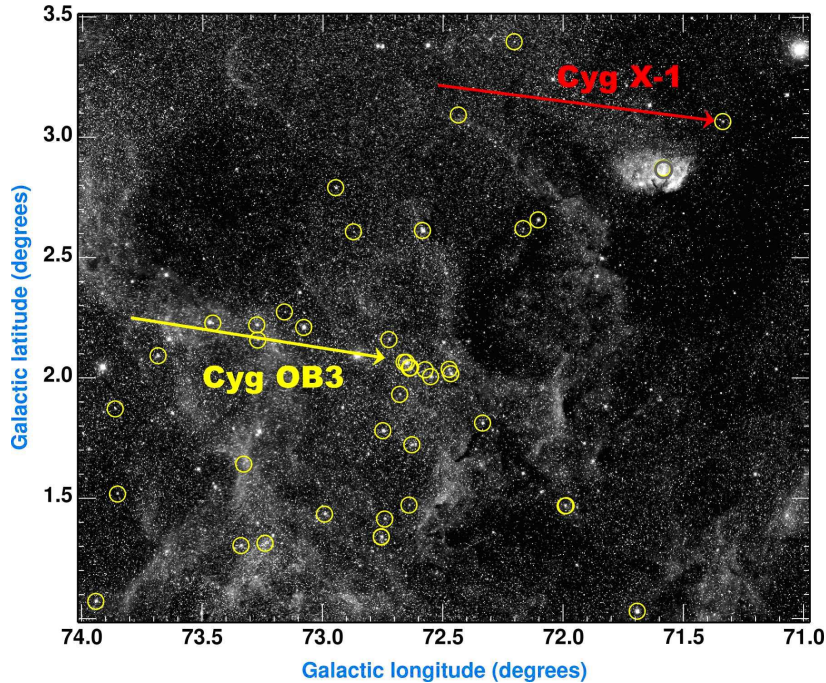} %{vbfig1_v2.eps}
\caption{The kinematics represented in this optical image of the sky
around the black hole x-ray binary Cygnus X-1 and the association of
massive stars Cygnus OB3 shows that Cygnus X-1 remained anchored at
its birth place in the association Cygnus OB3. The red arrow
represents the motion in the sky of the radio counterpart of Cygnus
X-1 for the past 0.5 million years. The yellow arrow the average
Hipparcos motion of the massive stars of Cygnus OB3 (circled in
yellow) for the past $0.5$ million years. Despite the different
observational techniques used to determine the proper motions,
Cygnus X-1 moves in the sky as Cygnus OB3. At a distance of 2 kpc,
the space velocity of Cygnus X-1 relative to that of Cygnus OB3 is
$< 9 \ {\rm km}/{\rm s}$, from which it is inferred that not more
than one solar mass could have been suddenly ejected during the
formation of the black hole in a natal supernova. From Mirabel, I.F.
\& Rodrigues (2003).} \label{mq}
\end{figure*}

The low mass x-ray binaries GRS 1915+105 and V404 Cyg that contain
black holes of $14$ and $12$ solar masses respectively have both
peculiar velocities relative to their environment of several tens of
km/s that lie on the plane of the Galaxy and are mostly in radial
direction towards the Galactic Centre. However, their velocity
components perpendicular to the Galactic plane are very small,
$10\pm4 \  {\rm km}/{\rm s}$ for GRS 1915+105 (Dhawan et al. 2007)
and $0.2\pm 3 {\rm km}/{\rm s}$ for V404 Cyg (Miller-Jones et al.
2009). The kinematics of pulsars show that natal kicks have no
preferential direction, and in this context the large peculiar
velocities on the plane of the Galaxy of these two low mass X-ray
binaries with black holes of more than $10$ solar masses are most
probably due to Galactic diffusion, rather than to energetic natal
kicks.

d) The theoretical expectation that very massive stars with high
metal content may end as neutron stars instead of black holes
(Meynet \& Maeder 2005) is consistent with observations of some
newly formed neutron stars. In fact, young neutron stars observed as
Soft Gamma Ray Repeaters (SGRs) and Anomalous X-ray Pulsars (AXPs)
are found in young clusters of massive stars. SGR 1806-20 (Mirabel
et al. 2000, Figer et al. 2005) and AXP CXOU J1647-45 are associated
with parent clusters of massive stars, the latter being inside the
young cluster Westerlund 1 (Muno et al. 2006). These clusters are in
the inner Galaxy and have metal contents larger than solar. Assuming
coeval formation of the most massive stars in the clusters and the
progenitors of the young neutron stars, lower limits of 40 and 50
solar masses for the masses of the progenitors of these neutron
stars are inferred (Figer et al. 2005, Muno et al. 2006).

e) Gamma Ray Bursts of long duration (LGRBs) mark the formation of
black holes by the collapse of massive stars. Although a fraction of
dark GRBs may require local extinction columns of $Av > 1 {\rm
mag}$, the majority of the hosts of LGRBs are faint, irregular
galaxies of limited chemical evolution (Le Floc'h et al. 2003,
Fruchter et al. 2006). GRB 060505 and GRB 060614 were observed
(Della Valle et al. 2006, Fynbo et al. 2006, Gal-Yam et al. 2006)
with no luminous SNe, and one possible explanation -among others
(Gehrels et al. 2006)- is black hole formation by direct collapse.

f) A number of supernovae, classified as core collapse Type II show
extremely low expansion velocity and an extraordinarily small amount
of $^{56}$Ni in the ejecta (Zampieri et al. 2003). These SNe are
under-energetic with respect to a typical Type II supernova and may
originate from the explosion of a massive progenitor in which the
rate of early infall of stellar material on the collapsed core is
large. Events of this type could form a black hole remnant, giving
rise to significant fallback with late-time accretion and relatively
small kicks. Recently it has been observed a low-energy
core-collapse supernova without hydrogen envelope (Valenti et al.
2009), suggesting a link of faint supernovae with long-duration
gamma-ray bursts.

g) There is increasing evidence for enhanced LGRB rate at $z > 3$,
as expected from the increase of specific star formation rate with
decreasing metallicity (Daigne et al. 2006, Kistler et al. 2008, Qin
et al. 2010).

\section{Feedback from accreting stellar black holes}
As in AGN, feedback from accreting stellar black holes is observed
in the form of energetic radiation, powerful jets, and massive
winds.

\textbf{Energetic radiation:} Super-Eddington radiation that last
relatively short times is observed in outbursts of black hole novae,
which have low-mass stellar donors (e.g. Nova Muscae, V404 Cyg).
BH-HMXBs can exhibit a somewhat steady form of super-Eddington
radiation of up to $10^{41} {\rm erg}/{\rm s}$. This is observed in
the extragalactic ULXs, which often have spectra that resemble the
very high (super-soft) state observed in some Galactic black hole
binaries (e.g. GRS 1915+105). The majority of ULXs exhibit a complex
curvature spectrum which can be modeled by a cool disk component
together with a power law which breaks above $3$ keV, probably due
to a cool, optically thick corona produced by super-Eddington
accretion flows (Gladstone et al. 2009).

On the other hand, the satellites Fermi and Agile detected gamma-ray
flares from two high mass X-ray binaries: from the microquasar
Cygnus X-3 and from Cygnus X-1. TeV flares may have been observed
with the Cherenkov telescope MAGIC. This high energy emission from
microquasars has been modeled in the context of jet leptonic and
hadronic models (Vila \& Romero 2010, Romero 2008).

\textbf{Powerful jets:} Steady large-scale jets have been observed
in several Galactic black hole binaries (e.g. 1E 1740.7-2942; GRS
1758-258; SS 433), which are viewed as small scale analogues of the
extragalactic FR II radio galaxies. The existence of powerful
super-Eddington jets of low radiation efficiency has been revealed
by the multiwavelength observations of the HMXB SS 433, where the
jets entrain atomic nuclei with velocities of $0.26 \ c$, have
mechanical energies $> 10^{39} {\rm erg}/{\rm s}$, and are capable
of blowing laterally the nebula W50 up to distances of tens of
parsecs. Large scale bow-shocked nebula produced by powerful dark
jets have been observed in the BH-HMXB Cygnus X-1 (Gallo et al.
2005), and more recently in the HMXB S26 in the galaxy NGC 7793
(Pakull et al. 2010). The mechanical energy injected by S26 is
$>10^{40}$ erg/s, showing that the overall energy injected by these
HMXB microquasars during their whole lifetime can be several orders
of magnitude that of the photonic and baryonic energy from a typical
core collapse supernova. Further evidence for powerful jets of low
radiation efficiency has been obtained by the observations in the
X-rays with Chandra of moving jets in the black hole binaries XTE
J1550-564 (Corbel et al. 2002) \& H1743-32 (Corbel et al. 2005).
These observations show in real time the formation of double-lobe
X-ray and radio lobes. The X-rays are produced by synchrotron
mechanism, implying that electrons are accelerated up to TeV
energies by shocks in the moving lobes at parsec distances from the
BHXRBs.

\textbf{Massive winds:} The connection between winds, jets and x-ray
emission has been observed with unprecedented detail in GRS 1915 105
(Neilsen et al. 2011).  The X-ray observations reveal ejections from
the inner disk in the form of jets followed by winds from the outer
disk that are massive enough to quench the jet and produces
transitions in the X-ray overall output. Non-relativistic massive
outflows in the form of winds of ionized and neutral gas have now
been observed in accreting black holes of all mass scales.

\section{Discussion and results}

\textbf{Ionization and thermal history of the Intergalactic Medium}
\vspace{1mm}

Early star-forming galaxies as the ones recently discovered with the
Hubble Space Telescope up to $z \sim 8$ (Bouwens et al. 2010), when
the universe was only $\sim 800$ millon years old, must have caused
the reionization of the intergalactic medium (IGM). However, it is
currently believed that the ultraviolet radiation from the massive
stars in those galaxies was enough to produce and keep over large
volumes of space most of the IGM reionized and heated to
temperatures of $\sim 10^4$ K (Roberston et al. 2010). This belief
resides in the assumption that the escape fraction into the IGM of
UV photons $f_{\rm esc} = 0.1-0.2$ as in galaxies at $z\sim 3$.
Furthermore, the recombination rate depends -besides on the hydrogen
density-, on the temperature of the IGM. Since the mean free path of
the X-rays in a neutral medium is much larger than that of the UVs
and X-rays are capable of producing multiple ionizations, Mirabel et
al. (2011) propose that the X-ray radiation from the large
population of BH-HMXBs in the firsts star burst galaxies ionized and
heated to temperatures of $\sim 10^4$ K the IGM at large distances in
the low density regions, bringing down the recombination rate, and
therefore keeping the whole IGM ionized.

The idea that black holes may have played a role in the reionization
era of the universe was developed previously by Madau et al. (2004).
Motivated by the early reports from the Wilkinson Microwave Probe
(WMAP) of a large optical depth to Thomson scattering that would
have implied a very early reionization epoch, it was proposed a
scenario where the universe was first reionized by intermediate
black holes at $z > 20$. However, a more accurate determination from
WMAP3 later lead to a revision of the optical depth downward (Page
et al. 2007), to a value consistent with reionization significantly
later, at the epoch when the first stars were form. On the other
hand, it has been shown (Alvarez et al. 2009, Milosavljevic et al.
2009) that feedback from accretion to solitary black holes
significantly affects further inflow and the consequent injection of
radiation and high energy particles to the surrounding medium.
Therefore, if X-rays played a role in the heating and reionization
of the Intergalactic Medium (IGM), in the context of the current
simulations of the first generations of massive stars (Krumholz et
al. 2009, Turk et al. 2009, Stacy et al. 2009), and the observations
of stellar black holes in the near and distance universe, the most
realistic alternative to quasi-radial, Bondi-like accretion would be
accretion to black hole stellar remnants in the first generations of
high mass binary stars, namely, BH-HMXBs.

On the other hand, current observational results indicate that
galaxies -and therefore stellar black holes- were formed before
supermassive black holes. In this context, heating of the IGM was
first caused and maintained by accreting stellar black holes, before
this role was taken over by AGN.

\vspace{3mm}
\textbf{Stellar black holes constrain the properties of dwarf galaxies}
\vspace{1mm}

The apparent disparity between the number of dwarf galaxies
predicted by the Cold Dark Matter model of the universe, with the
number of low mass galaxies observed so far in the halo of the
Galaxy is a subject of topical interest in cosmology. Power et al.
(2009) pointed out the possible implications of X-ray binaries in
primordial globular clusters for the reionization, and therefore,
galaxy formation at high redshifts.

As shown by Mirabel et al. (2011), once the IGM is heated to a
temperature of $10^4$ K, dark matter halos with masses below $10^9
M_\odot$  no longer can accrete IGM material because the temperature
of the infalling gas increases by an extra order of magnitude as its
density increases on its way into these galaxies. In that regime,
only gaseous halos with virial temperatures above $10^5$ K could
have accreted fresh IGM gas and converted it to stars. The census of
dwarf galaxy satellites of the Milky Way requires a related
suppression in the abundance of low-mass galaxies relative to
low-mass dark matter halos (Alvarez et al. 2009). \textit{The
thermal history of the IGM therefore has a direct impact on the
properties of the faintest galaxies at high redshifts as well as the
smallest dwarf galaxies in the local universe.}

It is interesting to note that black holes of different mass scales
have a role in galaxy formation. Feedback from supermassive black
holes halt star formation, quenching the unlimited mass growth of
massive galaxies (Fabian 2009). Feedback from stellar black holes in
HMXBs during the reionization epoch suppress the number of dwarf
galaxies with masses  $< 10^9 M_\odot$ . Therefore, BH-HMXBs in the
early universe are an important ingredient to reconcile the apparent
disparity between the observed number of dwarf galaxies in the
Galactic halo with the number of low mass galaxies predicted by the
Cold Dark Matter model of the universe.

%\begin{acknowledgements}
%\end{acknowledgements}

\begin{discussion}
\discuss{Pe'er}{Where would you draw the line between jet
(continuous) and ``blobs" ejection as seen in GRS 1915; is there
evidence for jet there at all?} \discuss{Mirabel}{In GRS 1915+105 we
have imaged with the VLBA a compact ``continuous" jet long of tens
of astronomical units and velocities $< 0.2 \, c$ associated to the
low-hard state. The sudden transitions from the low-hard state to
the high soft state mark the onset of discrete and bright
condensations or shocks that move away in the form of collimated
jets at apparent superluminal motions.}

\discuss{Kylafis}{In the slide that you showed about GRS 1915, the
optical should appear also. Was it detected?} \discuss{Mirabel}{No,
because GRS 1915 is on the Galactic plane at $> 6$ kpc and there are
about $30$ magnitudes of optical absorption along the line of
sight.}

\discuss{Sambruna}{In AGN (and GRS 1915) disk winds are observed
(see F. Tombesi contribution) that most likely contribute to feedbak
on large scales.} \discuss{Mirabel}{I fully agree. In fact, besides
the observations of ionized gas with Chandra reported in this
meeting by Neilsen et al., recent observations of Mrk 231 revealed
massive outflows of molecular gas. The ions moving at $0.26 \, c$ in
SS433 are from winds that have been entrained and accelerated by the
jets.}

\discuss{Kawai}{Half of GRBs are so-called ``dark" GRBs with no
optical counterpart. They may be formed in metal-rich environments.
Thus, low-metalicity is not an essential condition for GRB
formation.} \discuss{Mirabel}{In a recent study, Perley et al.
conclude that the source of obscuring dust is local to the vicinity
of the GRB progenitor and may be highly unevenly distributed within
the host galaxy. In fact, my former PhD student Emeric Le Floc'h did
not detect with Spitzer any of the host galaxies of dark GRBs, which
implies that these are not globally speaking, highly dusty galaxies.
The production of dust may occur rapidly in associations of massive
stars and the GRB could take place after other massive stars have
already enriched locally the interstellar medium.}

\discuss{Fendt}{You have been introducing the ``universal model for
jet-disk coupling" for BH. Why should it not work for neutron stars?
Central masses are similar, also the disks should be similar (\&
simirlaly relativistic). } \discuss{Mirabel}{It may work for neutron
stars, white dwarfs, and young stars. But I prefer to leave a more
detailed answer to this question to the speakers that will discuss
this issue later in the week.}

\discuss{Rodr\'\i guez}{To correct for these two failures of the
Cold Dark Matter model, do you need stellar-mass or supermassive
black holes, or both?} \discuss{Mirabel}{Both. Supermassive black
holes may account for the absence of masive cusps in the central
regions of galaxies. Stellar black holes in the early universe would
have heated the IGM to temperatures of about $10^4$ K which would
have impeded the formation of large numbers of low-mass galaxies by
baryonic accretion onto small haloes of dark matter.}

\discuss{Nishikawa}{What causes the time lag between infrared and
radio emission in GRS 1915+105?} \discuss{Mirabel}{The time delay is
due to the fact that in an adiabaticaly expanding plasma cloud, the
radiation becomes transparent to longer wavelenths at later times.}

\end{discussion}
\end{document}